\documentclass[12pt]{article}
\usepackage{amsmath,amssymb,bm,graphicx}
\usepackage{color} 

\begin{document}

\begin{flushright}
\end{flushright}

\begin{center}
{\Large{\bf Absence of  Majorana-Weyl fermions in d=4 and the theory of Majorana fermions}}
\end{center}
\vskip .5 truecm
\begin{center}
\bf { Kazuo Fujikawa }
\end{center}
\begin{center}
\vspace*{0.4cm} 
{\it {Center for Interdisciplinary Theoretical and Mathematical Sciences,\\
RIKEN, Wako 351-0198, Japan}
}
\end{center}
\makeatletter
\makeatother


\begin{abstract} 
It is customary to identify $\psi_{+}=\nu_{R} + C\overline{\nu_{R}}^{T}$ with a Majorana fermion on the basis of chirality changing charge conjugation $\tilde{C}: \nu_{R}\rightarrow  C\overline{\nu_{R}}^{T}$ and parity $\tilde{P}: \nu_{R}\rightarrow  i\gamma^{0}\nu_{R}$. The theorem on the absence of a Majorana-Weyl fermion in $d=4$ states $\tilde{C}\gamma_{5}\tilde{C}^{-1}= -\gamma_{5}$ with $\tilde{C}=C\gamma_{4}^{T}$, and thus the charge conjugation of the equivalent Majorana $\psi_{+}=(\frac{1+\gamma_{5}}{2})\nu_{R} + (\frac{1-\gamma_{5}}{2})C\overline{\nu_{R}}^{T}$ vanishes without subsidiary $\gamma_{5}\rightarrow - \gamma_{5}$, namely, not defined in field theory.
 To be consistent with the theorem, it is common to use  a doublet representation of chirality preserving charge conjugation $\hat{C}:\nu_{R,L}\rightarrow  C\overline{\nu_{L,R}}^{T}$ and parity  $\hat{P}: \nu_{R,L}\rightarrow i\gamma^{0}\nu_{L,R}$ in  theory containing both $\nu_{R,L}$.   In the type I seesaw model, the latter formulation is applicable but $\psi_{+}=\nu_{R} + C\overline{\nu_{R}}^{T}$ is not a Majorana fermion. An analogue of the Bogoliubov transformation converts $\psi_{\pm}=\nu_{R, L} \pm C\overline{\nu_{R, L}}^{T}$, which are obtained by a precise diagonalization of the seesaw model,  to Majorana fermions $\psi_{M_{1,2}}=(\psi \pm C\overline{\psi}^{T})/\sqrt{2}$ with a Dirac-type fermion $\psi$, as originally defined by Majorana.  A chiral projection $[(1+\gamma_{5})/2] \psi_{M_{1}}$ of a Majorana fermion is not a chiral fermion, which ensures the presence of the neutrino-less double beta decay.

\end{abstract}

\section{Introduction}
It is important to analyze a possible generalization of the Majorana fermion $\psi_{M}=(\psi\pm C\overline{\psi}^{T})/\sqrt{2}$ with a Dirac-type fermion $\psi$, as originally defined by Majorana \cite{Majorana},  to theories which contain the chiral operator $\gamma_{5}$.  To be specific, we are going to show that the formal construction 
\begin{eqnarray}\label{Majorana}
\psi_{+}=\nu_{R}+ C\overline{\nu_{R}}^{T},
\end{eqnarray}
which is very similar to the original construction of a Majorana fermion in terms of a Dirac fermion $\psi$, is not a Majorana fermion. The common use of a chirality changing C operation $\nu_{R}^{C}=C\overline{\nu_{R}}^{T}$ (i.e., the right-handed $\nu_{R}$ is transformed to the left-handed $C\overline{\nu_{R}}^{T}$) in the fermion \eqref{Majorana}, which is implied by the representation theory of Lorentz symmetry, is also a consequence of the theorem on the absence of a Majorana-Weyl fermion in d=4. The formal Majorana fermion \eqref{Majorana} based on the no-go theorem in d=4 is {\em indeterminate}; it sometimes behaves like a Majorana fermion but it in fact vanishes completely under the assumed chirality changing C symmetry. To my knowledge, this fact has not been discussed in the past, and it is shown in more detail later.  Another common argument for the Majorana \eqref{Majorana} is based on the relation $\psi_{+}=C\overline{\psi_{+}}^{T}$, which is the same as  the above chirality changing C operation by exchanging the first and the second terms, gives only a necessary condition; we have a continuous number of solutions for $\psi_{M}=C\overline{\psi_{M}}^{T}$ with $\psi_{\theta}=\cos\theta \nu_{R}+ \sin\theta\nu_{L}$ and $\Psi_{M}=\psi_{\theta}+C\overline{\psi_{\theta}}^{T}$. In comparison, the C symmetry of a Dirac fermion $\psi$ is uniquely specified.  Our past arguments against the Majorana fermion \eqref{Majorana} were based on the logical inconsistency \cite{Fujikawa2,Fujikawa-Tureanu 2019}, but in the present paper  it is  neatly explained as a consequence of the theorem on the absence of the Majorana-Weyl fermion in d=4.  The consistent Majorana fermion in d=4 is  given uniquely by the original expression of Majorana \cite{Majorana}. It is shown later  that this seemingly technical issue decides the presence or absence of the neutrino-less double beta decay in weak interactions. 

\section{Conventional  C, P and CP}  

  We now present the detailed analyses of the issues stated above. The representation with $\gamma_{5}$ diagonal is convenient when discussing Majorana fermions in field theory.
 The representation with $\gamma_{5}$ diagonal is defined by \cite{Schechter}, 
\begin{eqnarray}\label{conventional metric conventions}
\vec{\gamma}=\left(\begin{array}{cc}
            0&-i\vec{\sigma}\\
            i\vec{\sigma}&0
            \end{array}\right),
\gamma_{4}=\left(\begin{array}{cc}
            0&1\\
            1&0
            \end{array}\right),  
\gamma_{5}=\left(\begin{array}{cc}
            1&0\\
            0&-1
            \end{array}\right),  
C=\left(\begin{array}{cc}
            -\sigma_{2}&0\\
            0&\sigma_{2}
            \end{array}\right)
\end{eqnarray}  
and 
the 4-component Dirac-type field is parameterized without assuming a specific form of the Dirac equation by 
\begin{eqnarray}\label{4-component Dirac}
&&\psi(x) =\left(\begin{array}{c}
            \chi\\
            \sigma_{2}\phi^{\star}
            \end{array}\right) , \ \  
\psi^{C}(x) =C\overline{\psi}^{T}(x)=\left(\begin{array}{c}
            \phi\\
            \sigma_{2}\chi^{\star}
            \end{array}\right).     
 \end{eqnarray}  
 The fields with the parameterization \eqref{4-component Dirac} without specific Dirac equations are called Dirac-type fields in the present paper.
 The above C transformation implies in terms of two-component spinors
 \begin{eqnarray}\label{C of two-spinor}
  \hat{C}:\ \chi \rightarrow \phi,\ \ \phi \rightarrow \chi,
  \end{eqnarray}
 namely, in the chiral notation 
 \begin{eqnarray}
\hat{C}:\   \psi_{R}=\left(\begin{array}{c}
            \chi\\
            0
            \end{array}\right) \rightarrow                  
 C\overline{\psi_{L}}^{T}=\left(\begin{array}{c}
            \phi\\
            0
            \end{array}\right), \
 \psi_{L}=\left(\begin{array}{c}
            0\\
            \sigma_{2}\phi^{\star}
            \end{array}\right) \rightarrow                  
 C\overline{\psi_{R}}^{T}=\left(\begin{array}{c}
            0\\
            \sigma_{2}\chi^{\star}
            \end{array}\right) .
\end{eqnarray}
We thus tentatively abstract the C and P transformation laws of chiral fermions as        
\begin{eqnarray}\label{C and P}
\hat{C}&:& \ \nu_{L}(x)\rightarrow \nu^{C}_{L}(x)=C\overline{\nu_{R}}^{T}(x), \ \ \nu_{R}(x)\rightarrow \nu^{C}_{R}(x)= C\overline{\nu_{L}}^{T}(x),\nonumber\\
\hat{P}&:& \ \nu_{L}(x)\rightarrow\nu^{P}_{L}(x)= i\gamma_{4}\nu_{R}(t,-\vec{x}), \ \ \nu_{R}(x)\rightarrow \nu^{P}_{R}(x) =i\gamma_{4}\nu_{L}(t,-\vec{x}).
\end{eqnarray}
where the parity is defined by the chiral projections of $\psi(x)^{P}=i\gamma_{4}\psi(t,-\vec{x})$, and the combined CP is defined by 
\begin{eqnarray}\label{CP}
\hat{C}\hat{P}:\  \nu_{L}(x)\rightarrow\nu^{CP}_{L}(x)= i\gamma_{4}C\overline{\nu_{L}}^{T}(t,-\vec{x}), \ \ \nu_{R}(x)\rightarrow \nu^{CP}_{R}(x)= i\gamma_{4}C\overline{\nu_{R}}^{T}(t,-\vec{x}).
\end{eqnarray}
The above chirality preserving representation of the charge conjugation $\hat{C}$ with the doublet $(\nu_{R},\nu_{L})$ (a Dirac-type fermion) is consistent in field theory. Later we discuss a different chirality changing charge conjugation $\tilde{C}$ and a parity $\tilde{P}$ also, although they are rejected by the theorem on the absence of Majorana-Weyl fermions in d=4. 
 The extra $i$ for the parity (mirror symmetry) which is characteristic to the Majorana fermion, as was noted by Majorana himself (footnote 2 in the paper \cite{Majorana}), is incorporated. This extra $i$ is applied to all the chiral fermions but it does not influence the ordinary fermion number conserving terms. 
 
 We next recall the type I seesaw model \cite{Yanagida} (in the representation with $\gamma_{5}$ diagonal \cite{Schechter} from now on)
\begin{eqnarray}\label{Seesaw}
{\cal L}_{\nu}&=&
-\overline{\nu_{L}}(x)\gamma_{\mu}\partial_{\mu}\nu_{L}(x)
-\overline{\nu_{R}}(x)\gamma_{\mu}\partial_{\mu}\nu_{R}(x)\nonumber\\
&&- \{ \overline{\nu_{L}}m_{D}\nu_{R}(x) 
+(1/2)\nu^{T}_{R}(x)Cm_{R}^{\dagger}\nu_{R}(x)+h.c.\}
\end{eqnarray}
which is obtained by adding a $3\times 3$ complex symmetric mass term (a dimension 3 operator) to the gauge singlet $\nu_{R}$, in the model of massive Dirac neutrinos in an extension of the Standard Model.  We choose $m_{D}$ as a real diagonalized $3\times 3$ matrix.

 In the CPT theorem, we assign definite transformation laws  C, P and T to the irreducible representations of the proper Lorentz transformation, but in practice only CP is important. In contrast, the charge conjugation becomes crucial in the analysis of a Majorana fermion.
The above choice of discrete symmetry transformation laws \eqref{C and P} and \eqref{CP} is consistent since the action \eqref{Seesaw} is CP invariant for the real mass matrix $m_{R}$. (If $\nu_{R}=0$, we understand that C and P are not defined separately for $\nu_{L}$ and only the combined CP is defined.) Our task is to construct Majorana fermions with a proper C symmetry  in a lepton-number violating theory of chiral fermions \eqref{Seesaw}. We exploit an analogy of \eqref{Seesaw} with the BCS Lagrangian when one replaces the continuous electromagnetic symmetry with the discrete C symmetry \cite{comment}.

The mass term in \eqref{Seesaw} is diagonalized by writing it as 
\begin{eqnarray}\label{mass term}
(-2){\cal L}_{mass}=
\left(\begin{array}{cc}
            \overline{\nu_{R}}&\overline{\nu_{R}^{C}}
            \end{array}\right)
\left(\begin{array}{cc}
            m_{R}& m_{D}\\
            m_{D}^{T}&0
            \end{array}\right)
            \left(\begin{array}{c}
            \nu_{L}^{C}\\
            \nu_{L}
            \end{array}\right) +h.c.,
\end{eqnarray}
where 
$\nu_{L}^{C} = C\overline{\nu_{R}}^T$ and $\nu_{R}^{C} = C\overline{\nu_{L}}^T$.  
Since the above mass matrix is complex and symmetric, one can diagonalize it precisely by a $6 \times 6$ unitary $U$ (Autonne--Takagi factorization \cite{Autonne--Takagi})  as
\begin{eqnarray}\label{orthogonal}
            U^{T}
            \left(\begin{array}{cc}
            m_{R}& m_{D}\\
            m_{D}^{T}& 0
            \end{array}\right)
            U
            =\left(\begin{array}{cc}
            M_{1}&0\\
            0&-M_{2}
            \end{array}\right)    ,        
\end{eqnarray}
where $M_{1}$ and $M_{2}$ are $3\times 3$ real diagonal matrices. The signature of $M_{2}$ is chosen to make $M_{2}$ real and positive for an explicitly solvable case of a single flavor.

We can thus rewrite the seesaw Lagrangian \eqref{Seesaw} using the canonical transformation, which does not change the form of the kinetic  terms in \eqref{Seesaw} and thus anti-commutators,  
\begin{eqnarray} \label{variable-change}          
            &&\left(\begin{array}{c}
            \nu_{L}^{C}\\
            \nu_{L}
            \end{array}\right)
            = U \left(\begin{array}{c}
            \tilde{\nu}_{L}^{C}\\
            \tilde{\nu}_{L}
            \end{array}\right)
           ,\ \ \ \ 
            \left(\begin{array}{c}
            \nu_{R}\\
            \nu_{R}^{C}
            \end{array}\right)
            = U^{\star} 
            \left(\begin{array}{c}
            \tilde{\nu}_{R}\\
            \tilde{\nu}_{R}^{C}
            \end{array}\right),         
\end{eqnarray}
and using the mass eigenvalues $M_{1}$ and $M_{2}$.  We are using $U$ and $U^{T}$ instead of $U$ and $U^{\dagger}$, with particles and anti-particles combined in 6-dimensional flavor space. Our transformation should be regarded as a canonical transformation defined at a specific time $t=0$, for example.  See also an analogous Bogoliubov transformation in the model of Nambu-Jona-Lasinio \cite{Nambu}.

We thus obtain  \cite{Fujikawa-Tureanu 2019, Fujikawa-Tureanu-2024} (in the representation with $\gamma_{5}$ diagonal \cite{Schechter})
\begin{eqnarray}\label{exact-solution}
{\cal L}_{\nu}
 &=&-(1/2)\{\overline{\psi_{+}}[\gamma_{\mu}\partial_{\mu}\ + M_{1}]\psi_{+} +\overline{\psi_{-}}[\gamma_{\mu}\partial_{\mu}\  + M_{2}]\psi_{-}\} ,
\end{eqnarray}
 with (by suppressing the tilde-symbol of $\tilde{\nu}$)
\begin{eqnarray}\label{pseudoMajorana}
\psi_{+}(x)&=& \nu_{R}+ \nu^{C}_{L}= \nu_{R}+ C\overline{\nu_{R}}^{T},\nonumber\\
\psi_{-}(x)&=&\nu_{L}- \nu^{C}_{R}=\nu_{L}- C\overline{\nu_{L}}^{T}.
\end{eqnarray}
The Lagrangian \eqref{exact-solution} with real mass parameters is CP invariant, and the possible CP breaking contained in the matrix $U$ is transferred to the  Pontecorvo-Maki-Nakagawa-Sakata (PMNS) weak mixing matrix \cite{Fujikawa-Tureanu-2024}.  These two fermions \eqref {pseudoMajorana} satisfy the necessary conditions of Majorana fermions 
\begin{eqnarray}\label{Formal Majorana}
\psi_{+}(x) =C\overline{\psi_{+}}^{T}(x), \ \ \psi_{-}(x) =-C\overline{\psi_{-}}^{T}(x)
\end{eqnarray}  
but the operator basis of this Majorana property using the C transformation laws of each constituent chiral fermions in \eqref{C and P} is missing.   
For example, \eqref{C and P} implies
\begin{eqnarray}\label{non-Majorana}
\psi_{+}^{C}=\nu_{L}+ C\overline{\nu_{L}}^{T},\ \ \psi_{-}^{C}=-\nu_{R}+ C\overline{\nu_{R}}^{T},
\end{eqnarray}
although CP in \eqref{CP} is natural
\begin{eqnarray}
\psi_{+}^{CP}(x)= i\gamma_{4}\psi_{+}(t,-\vec{x}), \ \ \psi_{-}^{CP}(x)=- i\gamma_{4}\psi_{-}(t,-\vec{x}).
\end{eqnarray}
If one should assume that $\psi_{+}$ in \eqref{pseudoMajorana} is a Majorana fermion, one would have 
\begin{eqnarray}\label{inconsistency}
[(1+\gamma_{5})/2]\psi_{+} = \nu_{R},
\end{eqnarray}
of which left-hand side is C invariant while the right-hand side is not in general.

\section{Chirality changing C  and P }

In the past (and presently also), a different definition of charge conjugation has been widely used to justify the interpretation of the fermion $\psi_{+}$ in \eqref{pseudoMajorana} as a Majorana fermion. (In the following we mainly use $\psi_{+}$ as a representative example of a fermion.) 
The action for fermions in \eqref{exact-solution} is written  by noting \eqref{pseudoMajorana} as 
\begin{eqnarray}\label{exact-solution2}
&&\int d^{4}x(-1/2)\{\overline{\psi_{+}}[\gamma_{\mu}\partial_{\mu}\ + M_{1}]\psi_{+} +\overline{\psi_{-}}[\gamma_{\mu}\partial_{\mu}\  + M_{2}]\psi_{-}\} , \nonumber\\
 &=&\int d^{4}x \{ -\overline{\nu_{R}}(x)\gamma_{\mu}\partial_{\mu}\nu_{R}(x)
-[(1/2)\nu^{T}_{R}(x)CM_{1}\nu_{R}(x)+h.c.]\}\nonumber\\
&&\hspace {1cm} +\{ -\overline{\nu_{L}}(x)\gamma_{\mu}\partial_{\mu}\nu_{L}(x)
+[(1/2)\nu^{T}_{L}(x)CM_{2}\nu_{L}(x)+h.c.]\}.
\end{eqnarray}
The commonly chosen charge conjugation operation of $\psi_{\pm}(x)$ in \eqref{pseudoMajorana}  in the literature \cite{Bilenky} is 
  the chirality changing (i.e., the right-handed $\nu_{R}$ is transformed to the left-handed $C\overline{\nu_{R}}^{T}$ and vice versa, which was called {\em pseudo-C symmetry}  in \cite{Fujikawa-Tureanu 2019})  in place of \eqref{C and P},
\begin{eqnarray}\label{pseudo-C 2}
\tilde{C}:\ \ &&\nu_{R}\rightarrow \nu_{R}^{\tilde{C}}=C\overline{\nu_{R}}^{T},\ \ \ \ (\nu_{R}^{\tilde{C}})^{\tilde{C}} = (C\overline{\nu_{R}}^{T})^{\tilde{C}} \rightarrow \nu_{R}, \nonumber\\
&&\nu_{L}\rightarrow \nu_{L}^{\tilde{C}}=C\overline{\nu_{L}}^{T},\ \ \ \ (\nu_{L}^{\tilde{C}})^{\tilde{C}} = (C\overline{\nu_{L}}^{T})^{\tilde{C}} \rightarrow \nu_{L}
\end{eqnarray}
under which $\psi_{\pm}\rightarrow \pm \psi_{\pm}$ in \eqref{pseudoMajorana} and thus the action  \eqref{exact-solution2} is formally invariant. Note that this pseudo-C symmetry has been taken as the  true 
C operation in the literature \cite{Bilenky}.  The chirality changing parity in this scheme is defined by 
\begin{eqnarray}\label{P}
\tilde{P}&:& \ \nu_{L}(x)\rightarrow\nu^{\tilde{P}}_{L}(x)= i\gamma_{4}\nu_{L}(t,-\vec{x}), \ \ \nu_{R}(x)\rightarrow \nu^{\tilde{P}}_{R}(x) =i\gamma_{4}\nu_{R}(t,-\vec{x})
\end{eqnarray}
to reproduce $\tilde{C}\tilde{P}=CP$ in \eqref{CP} which appears to be reasonable.

One can confirm that the action \eqref{exact-solution2} is  invariant under the  pseudo-C transformation,
but the same action \eqref{exact-solution2} after using the {\em identities} 
$\nu_{R}=(\frac{1+\gamma_{5}}{2}) \nu_{R}$ and $\nu_{L}=(\frac{1-\gamma_{5}}{2}) \nu_{L}$,
\begin{eqnarray}\label{Inconsistency}
&&\int d^{4}x \{ -\overline{\nu_{R}}(x)\gamma_{\mu}\partial_{\mu}(\frac{1+\gamma_{5}}{2})\nu_{R}(x)
-[(1/2)\nu^{T}_{R}(x)CM_{1}(\frac{1+\gamma_{5}}{2})\nu_{R}(x)+h.c.]\}\nonumber\\
&&\hspace {1cm} +\{ -\overline{\nu_{L}}(x)\gamma_{\mu}\partial_{\mu}(\frac{1-\gamma_{5}}{2}) \nu_{L}(x)
+[(1/2)\nu^{T}_{L}(x)CM_{2}(\frac{1-\gamma_{5}}{2}) \nu_{L}(x)+h.c.]\}, \nonumber\\
\end{eqnarray}
exactly vanishes under the same pseudo-C in \eqref{pseudo-C 2}.  Also, both of 
\begin{eqnarray}\label{fields under non-commuting symmetry}
\psi_{+}=(\frac{1+ \gamma_{5}}{2})\nu_{R}+ (\frac{1- \gamma_{5}}{2})C\overline{\nu_{R}}^{T}, \ \ \psi_{-}=(\frac{1-\gamma_{5}}{2})\nu_{L}- (\frac{1+\gamma_{5}}{2})C\overline{\nu_{L}}^{T}
\end{eqnarray}
  vanish under the pseudo-C transformation \eqref{pseudo-C 2}. The substitution rules, which imply the universal validity of symmetry laws for the different ways of writing of the equivalent  Lagrangian (or field variables),  fail for chirality changing transformations.  In contrast, the CP transformation in the present scheme, which is chosen as the same as in the conventional scheme \eqref{CP} (i.e., chirality preserving), 
 satisfies all the normal properties including the substitution rules in \eqref{exact-solution2},  \eqref{Inconsistency} and \eqref{fields under non-commuting symmetry}.
 (The chirality changing C operation in \eqref{pseudo-C 2} is also shown to lead to inconsistencies in the two-component formalism, as is readily confirmed.)

This fact, namely, the pseudo-C cannot be used as a charge conjugation in  conventional field theory, 
has been emphasized in \cite{Fujikawa-Tureanu 2019, Fujikawa-Tureanu-2024}, although the pseudo-C implies formally the fermions in \eqref{pseudoMajorana} to be  Majorana fermions. Besides, the parity $\tilde{P}$ \eqref{P} is not a mirror symmetry in the conventional sense.   Also, the puzzling relation  of  the supposedly Majorana fermion \eqref{pseudoMajorana} as shown in \eqref{inconsistency} remains.

These puzzling properties are understood, although not resolved, if one recalls the theorem which states that the Majorana-Weyl fermions do not exist in d=4. This theorem is characterized by \cite{Ohta}
\begin{eqnarray}
\tilde{C}\gamma_{5}\tilde{C}^{-1}=-\gamma_{5}
\end{eqnarray}
in d=4, namely, a basic charge conjugation operator in the representation theory of Lorentz symmetry (for a singlet representation of a chiral fermion) cannot maintain the chirality.   The chirality changing pseudo-C corresponds to such an operator; in our notation the operator $\tilde{C}$ is given by $C\gamma_{4}^{T}$, as is explicitly confirmed. The chirality changing transformation itself is a manifestation of the no-go theorem in d=4, and the chirality changing  parity \eqref{P} also gives  $\tilde{P}\gamma_{5}\tilde{P}^{-1}=-\gamma_{5}$. To simulate the ordinary charge conjugation  by  the pseudo-C in the framework of conventional field theory, the c-number $\gamma_{5}$ needs to be transformed as if it were formally a field variable. Precisely speaking,  the theorem in the present context is understood as showing that one cannot use the properties of both a Majorana fermion and a chiral fermion simultaneously in d=4. In contrast, the chirality preserving $\hat{C}$ and $\hat{P}$ in \eqref{C and P} are not constrained by the theorem.  

The no-go theorem suggests that simultaneously with the pseudo-C operation as a substitution rule, one needs to transform  
\begin{eqnarray}\label{extra transformation}
\gamma_{5}\rightarrow -\gamma_{5},
\end{eqnarray}
everywhere in the Lagrangian (or field variables),  then \eqref{Inconsistency} and \eqref{fields under non-commuting symmetry} do not vanish although we operate outside the conventional field theory. Also the puzzling relation \eqref{inconsistency} is transformed to
\begin{eqnarray}\label{puzzling relation 2}
 [(1-\gamma_{5})/2]\psi_{+} = C\overline{\nu_{R}}^{T}
\end{eqnarray}
by the pseudo-C transformation supplemented by $\gamma_{5}\rightarrow -\gamma_{5}$. Now \eqref{inconsistency} is formally consistent, although it is still mysterious from a view point of conventional field theory that  a chiral field (chiral projection of a Dirac field) is a chiral projection of a Majorana field (real part of a Dirac field). This causes a problem in the analysis of the neutrino-less double beta decay in weak interactions where the chiral projection of a Majorana fermion appears, as is shown later.

Also, the kinetic part of a right-handed chiral fermion in \eqref{exact-solution2}
\begin{eqnarray}
\int d^{4}x \{-\overline{\nu_{R}}(x)\gamma_{\mu}\partial_{\mu}\nu_{R}(x)\} =
\int d^{4}x \{-\overline{\nu_{R}}(x)\gamma_{\mu}\partial_{\mu}(1+\gamma_{5})\nu_{R}(x)\}
\end{eqnarray}
is invariant under the pseudo-C symmetry \eqref{pseudo-C 2} and, separately, under $\tilde{P}$ \eqref{P} also, when combined with \eqref{extra transformation}; this is different from  a transformation property of a Weyl-type fermion in  conventional field theory \cite{Fujikawa-Tureanu 2019}. 
 The chirality changing C itself is a result of the no-go theorem and it leads to {\em indeterminate} physical predictions; it sometimes leads to a Majorana fermion in  \eqref{pseudoMajorana} formally but it actually leads to  vanishing results \eqref{fields under non-commuting symmetry}. 
We conclude that the chirality changing C is not used to identify the fermions  in \eqref{pseudoMajorana} to be  Majorana fermions.

Our choice of the chirality preserving C in \eqref{C  and P} is based on the doublet representation of chiral fermions $(\nu_{R}, \nu_{L})$ (namely, a Dirac-type fermion)
and it is consistent in conventional Lagrangian field theory. To realize Majorana fermions consistently, however, one needs to go first to a Dirac-type fermion by some means and then to a pair of Majorana fermions, in conformity with the original definition of Majorana fermions from a Dirac fermion \cite{Majorana}.

 \section{Bogoliubov-type transformation in seesaw model} 

So far in section 2, we have shown that the conventional C and P in field theory \eqref{C and P} lead  the exact solutions \eqref{pseudoMajorana} of the seesaw model to the fermions which are not Majorana fermions \eqref{non-Majorana}. In section 3, the commonly used chirality changing C in \eqref{pseudo-C 2} is shown to cause  the vanishing action in \eqref{Inconsistency} due to the theorem on the absence of the Majorana-Weyl fermion in d=4. Thus the exact solutions of the Type I seesaw model \eqref{Seesaw} have no sensible physical interpretations in the framework of conventional field theory.  
 
As a way to give a consistent meaning to $\psi_{\pm}(x)$ in \eqref{pseudoMajorana} in conventional field theory, it has been  shown in \cite{ Pauli, Fujikawa-Tureanu-2024} that an additional application of a  generalized Pauli-Gursey canonical transformation, which has the same form as a canonical transformation in \eqref{variable-change},  with a specific (CP preserving) $6\times 6$ orthogonal matrix 
\begin{eqnarray}\label{Pauli-Gursey}
O=\frac{1}{\sqrt{2}} \left(\begin{array}{cc}
            1&1\\
            -1&1
            \end{array}\right)
\end{eqnarray}
is useful. An arbitrary canonical mixing  of neutrinos  and  anti-neutrinos with a possible change of the definition of the vacuum is called a generalized Pauli-Gursey transformation (an analogue of the Bogoliubov transformation) in \cite{Pauli}. It gives (at $t=0$, for example), by suppressing the tilde symbol of $\nu$ variables in \eqref{variable-change},
 \begin{eqnarray} \label{Pauli--Gursey0}          
            &&\left(\begin{array}{c}
            \nu_{L}^{C}\\
            \nu_{L}
            \end{array}\right)
            = O \left(\begin{array}{c}
            N_{L}^{C}\\
            N_{L}
            \end{array}\right) =\frac{1}{\sqrt{2}}\left(\begin{array}{c}
            N_{L}^{C}+N_{L}\\
            - N_{L}^{C} +N_{L}
            \end{array}\right) 
           ,\nonumber\\ 
           &&  \left(\begin{array}{c}
            \nu_{R}\\
            \nu_{R}^{C}
            \end{array}\right)
            = O 
            \left(\begin{array}{c}
            N_{R}\\
            N_{R}^{C}
            \end{array}\right)=\frac{1}{\sqrt{2}}\left(\begin{array}{c}
            N_{R}^{C}+N_{R}\\
            - N_{R}^{C} +N_{R}
            \end{array}\right)
\end{eqnarray}
without changing mass parameters in \eqref{exact-solution}. To be explicit, we have the transformation
\begin{eqnarray}\label{Pauli-Gursey2}
&&\psi_{+}=\nu_{R}+C\overline{\nu_{R}}^{T}\rightarrow  \frac{1}{\sqrt{2}} (N_{R} +N_{R}^{C})+\frac{1}{\sqrt{2}}(N_{L} +N_{L}^{C})\equiv \psi_{M_{1}},\nonumber\\
&&\psi_{-}=\nu_{L}-C\overline{\nu_{L}}^{T}\rightarrow
\frac{1}{\sqrt{2}}(N_{L} -N_{L}^{C})-\frac{1}{\sqrt{2}}(N_{R}^{C}-N_{R})\equiv \psi_{M_{2}}.
\end{eqnarray}
with Dirac-type field $N(x)$ without changing the mass parameters in \eqref{exact-solution}. The appearance of a canonical transformation at this point may not be abrupt since we have already used a canonical transformation in \eqref{variable-change}.

In passing, when one defines using a Dirac-type field $\psi$
\begin{eqnarray}\label{conventional Majorana}
\psi_{M_{1}}=\frac{1}{\sqrt{2}}(\psi+C\overline{\psi}^{T}), \ \
\psi_{M_{2}}=\frac{1}{\sqrt{2}}(\psi-C\overline{\psi}^{T}),  
\end{eqnarray}                     
they satisfy the conditions of the conventional Majorana fermions.

\begin{eqnarray}\label{Majorana condition 2}
\psi_{M_{1,2}}^{C}=\pm \psi_{M_{1,2}},\ \ \ \psi_{M_{1}}=C\overline{\psi_{M_{1}}}^{T}, \ \  \psi_{M_{2}}=-C\overline{\psi_{M_{2}}}^{T}
\end{eqnarray}
using  the C transformation laws of the Dirac-type field $\psi$. In the definition of $\psi_{M_{2}}$ in \eqref{conventional Majorana}, we adopted $\psi_{M_{2}}(x)^{C}=C\overline{\psi_{M_{2}}(x)}^{T} =-\psi_{M_{2}}(x)$ which satisfies $\{\psi_{M_{2}}(x)^{C}\}^{C}=\psi_{M_{2}}(x)$.
This construction \eqref{conventional Majorana} is essentially the original definition of Majorana \cite{Majorana}. A difference is that we do not have the Dirac equation in the present problem.  But the parameterization of the Dirac-type field \eqref{4-component Dirac} is freely chosen and the charge conjugation (and parity) transformation laws \eqref{C and P} remain valid.  We thus use \eqref{conventional Majorana} and \eqref{Majorana condition 2} in the present context also by regarding those relations valid at a specific time $t=0$, for example.

Coming back to \eqref{Pauli--Gursey0}, the generalized Pauli-Gursey transformation gives
\begin{eqnarray}\label{True Majorana}
&&\psi_{+}\rightarrow \psi_{M_{1}}=\frac{1}{\sqrt{2}}\left(\begin{array}{c}
            N+N^{C}
            \end{array}\right) ,\nonumber\\
 &&\psi_{-}\rightarrow \psi_{M_{2}}=\frac{1}{\sqrt{2}}\left(\begin{array}{c}
            N-N^{C}
            \end{array}\right)         
\end{eqnarray}
without changing anti-commutation relations and mass parameters. Namely, we have converted the seesaw Lagrangian \eqref{exact-solution} (originally defined in \eqref{Seesaw}) to 
\begin{eqnarray}\label{exact-solution 2}
{\cal L}_{\nu}
 &=&-(1/2)\{\overline{\psi_{M_{1}}}[\gamma_{\mu}\partial_{\mu}\ + M_{1}]\psi_{M_{1}} +\overline{\psi_{M_{2}}}[\gamma_{\mu}\partial_{\mu}\  + M_{2}]\psi_{M_{2}}\} 
\end{eqnarray}
 containing only Dirac-type fermions $N(x)$ without chiral $\gamma_{5}$ by a canonical transformation. We thus have a consistent formulation of Majorana fermions by preserving \eqref{C and P}  \cite{Fujikawa-Tureanu-2024}. The parity transformations 
are determined by \eqref{C and P} to be $\psi_{M_{1}}\rightarrow i\gamma_{4}\psi_{M_{1}}(t,-\vec{x})$ and $\psi_{M_{2}}\rightarrow i\gamma_{4}\psi_{M_{2}}(t,-\vec{x})$. Note $P^{2}=-1$, as required \cite{Majorana}. The combined CP is thus naturally defined.

As for the two-component spinor formalism, one may start with
\begin{eqnarray}\label{Two-component spinor2}
&&\psi_{M_{1}}=\frac{1}{\sqrt{2}}\left(\begin{array}{c}
            N+N^{C}
            \end{array}\right)=\left(\begin{array}{c}
            \rho_{1}\\
            \sigma_{2}\rho_{1}^{\star}
            \end{array}\right) ,\nonumber\\
 &&\psi_{M_{2}}
            =\frac{1}{\sqrt{2}}\left(\begin{array}{c}
            N-N^{C}
            \end{array}\right) =\left(\begin{array}{c}
            \rho_{2}\\
            -\sigma_{2}\rho_{2}^{\star}
            \end{array}\right),         
\end{eqnarray}
with (see \eqref{4-component Dirac} for the parameterization of a Dirac-type field $N$) 
\begin{eqnarray}\label{ordinary transformation2}
\rho_{1}=\frac{1}{\sqrt{2}}(\chi+\phi),\ \
\rho_{2}=\frac{1}{\sqrt{2}}(\chi-\phi).
\end{eqnarray}
The C transformation properties are 
\begin{eqnarray}\label{C transformation2}
\hat{C}:\ \ \rho_{1}\rightarrow \rho_{1},\ \
\rho_{2}\rightarrow -\rho_{2},
\end{eqnarray}
namely, this change of the C transformation laws of the 2-component spinors $\rho_{1}$ and $\rho_{2}$  from \eqref{C and P} is regarded to be a result of the specific generalized Pauli-Gursey canonical transformation (a transformation from chiral fermions to a theory without chiral $\gamma_{5}$). Besides, the chiral projection of Majorana fermions is not expected to generate the conventional chiral fermions in \eqref{Two-component spinor2}; in fact, the C transformation of the relation $[(1+\gamma_{5})/2]\psi_{M_{1}}=\rho_{1}$ combined with \eqref{ordinary transformation2}, for example, is  now consistent.

The seesaw Lagrangian  \eqref{exact-solution} is then written by replacing $\psi_{+}\rightarrow \psi_{M_{1}}$ and $\psi_{-}\rightarrow \psi_{M_{2}}$ (with $\sigma_{\mu}= (\bar{\sigma},-i)$)
 \begin{eqnarray}\label{two component action}
 {\cal L}_{\nu}&=&-i\rho_{1}^{\dagger}\sigma_{\mu}\partial_{\mu}\rho_{1} -\frac{1}{2}M_{1}\rho_{1}^{T}\sigma_{2}\rho_{1}-\frac{1}{2}M_{1}\rho_{1}^{\dagger}\sigma_{2}\rho_{1}^{\star}\nonumber\\
&& +i\rho_{2}^{\dagger}\sigma_{\mu}\partial_{\mu}\rho_{2} -\frac{1}{2}M_{2}\rho_{2}^{T}\sigma_{2}\rho_{2}-\frac{1}{2}M_{2}\rho_{2}^{\dagger}\sigma_{2}\rho_{2}^{\star},
 \end{eqnarray}
 which in turn gives rise to the two-component spinor  equations,
 \begin{eqnarray}
 -i\sigma_{\mu}\partial_{\mu}\rho_{1}-M_{1}\sigma_{2}\rho_{1}^{\star}=0, \ \ \ i\sigma_{\mu}\partial_{\mu}\rho_{2}-M_{2}\sigma_{2}\rho_{2}^{\star}=0.
 \end{eqnarray}
The symmetries of the fields \eqref{ordinary transformation2} and  the above Lagrangian are
\begin{eqnarray}\label{discrete symmetry}
  \hat{C}&:&\ \ \rho_{1}\rightarrow \rho_{1},\ \ \rho_{2}\rightarrow -\rho_{2},\nonumber\\
 \hat{P}&:&\ \ \rho_{1}\rightarrow i\sigma_{2}\rho_{1}^{\star}(t,-\vec{x}), \ \ \sigma_{2}\rho_{1}^{\star}\rightarrow i\rho_{1}(t,-\vec{x}), \nonumber\\
&& \ \ \rho_{2}\rightarrow -i\sigma_{2}\rho_{2}^{\star}(t,-\vec{x}), \ \ -\sigma_{2}\rho_{2}^{\star}\rightarrow i\rho_{2}(t,-\vec{x})
\end{eqnarray}
 where the parity transformations correspond to $\psi_{M_{1}}\rightarrow i\gamma_{4}\psi_{M_{1}}(t,-\vec{x})$ and $\psi_{M_{2}}\rightarrow i\gamma_{4}\psi_{M_{2}}(t,-\vec{x})$. Note $P^{2}=-1$, as required \cite{Majorana}.  Our view is that the original construction by Majorana \cite{Majorana} is the only consistent way to construct the Majorana fermion in field theory in d=4 defined by the charge conjugation \eqref{C and P}. The above two-component formalism consists of two 2-component spinors as in \eqref{ordinary transformation2}; the consistent scheme of Majorana \cite{Majorana} uses the two 2-spinors not a single 2-spinor in each two-component  Lagrangian \eqref{two component action}. 
 
\section{Discussions} 

The Type I seesaw model is exactly diagonalized by a canonical transformation $U$ in \eqref{variable-change}, and the solution is given by $\psi_{+}=\nu_{R}+ C\overline{\nu_{R}}^{T}$, for example. 
All the complications of our past analyses \cite{Fujikawa-Tureanu 2019, Fujikawa-Tureanu-2024} arise from the failure of the substitution rules in the  field theory of $\psi_{+}$ (i.e., the transformation laws depend on the ways of writing of the equivalent Lagrangian) using the chirality changing C symmetry, which was called pseudo-C symmetry. This failure of the substitution rules is dictated by the theorem on the absence of the Majorana-Weyl fermion in d=4. To maintain the conventional CP, the parity also becomes chirality changing and exhibits unconventional properties including the failure of the substitution rules. A relation $[(1+\gamma_{5})/2]\psi_{+} = \nu_{R}$ in \eqref{inconsistency} is mysterious since a chiral fermion is  given  by a chiral projection of a Majorana fermion $\psi_{+}$, which influences the analysis of weak interactions such as a double $\beta$ decay (in the case of $\psi_{-}$). To be explicit, 
\begin{eqnarray}
\langle 0|T^{\star} (\frac{1-\gamma_{5}}{2})\psi_{M}(x) (\frac{1-\gamma_{5}}{2})\psi_{M}(y)|0\rangle\neq 0
\end{eqnarray}
 is a condition of the neutrino-less double beta decay for a Majorana neutrino $\psi_{M}$.  If one should identify $\psi_{-}(x)=\nu_{L}- C\overline{\nu_{L}}^{T}$ in \eqref{pseudoMajorana} to be a Majorana fermion using the chirality changing C,  one would have
\begin{eqnarray}\label{Double beta}
\langle 0|T^{\star} (\frac{1-\gamma_{5}}{2})\psi_{-}(x) (\frac{1-\gamma_{5}}{2})\psi_{-}(y)|0\rangle&=&(\frac{1-\gamma_{5}}{2})\langle 0|T^{\star} \nu_{L}(x) \nu_{L}(y)|0\rangle\\
&=&(\frac{1-\gamma_{5}}{2})\langle 0|T^{\star} C\overline{\nu_{L}}^{T}(x) C\overline{\nu_{L}}^{T}(y)|0\rangle =0\nonumber
\end{eqnarray}
 where we used $(\frac{1-\gamma_{5}}{2})\psi_{-}(x)=(\frac{1-\gamma_{5}}{2})\nu_{L}(x)$ and $(\frac{1-\gamma_{5}}{2})\psi_{-}(y)=\nu_{L}(y)$ in the first line and the assumed invariance of the vacuum $|0\rangle$ under the chirality changing C (pseudo-C) in the second line; $(\frac{1-\gamma_{5}}{2})$ in the second line of the right-hand side of \eqref{Double beta} acts on the right-handed $C\overline{\nu_{L}}^{T}(x)$. This vanishing of the second line was noted in \cite{Fujikawa1}. If one should use the first expression of the action \eqref{exact-solution2} in the starting expression of the  above \eqref{Double beta}, one would conclude the ordinary neutrino-less double beta decay, as is predicted by the Majorana fermion $\psi_{M_{2}}$ in \eqref{exact-solution 2}. The chirality changing charge conjugation based on  the theorem on the absence of the Majorana-Weyl fermion in d=4, which identifies $\psi_{-}$ in \eqref{pseudoMajorana} formally with the Majorana neutrino, leads to {\em indeterminate} conclusions. 
 One may conclude that a consistent field theoretical description of $\psi_{+}=\nu_{R}+ C\overline{\nu_{R}}^{T}$ as a Majorana fermion  is absent.

We have shown \cite{Pauli, Fujikawa-Tureanu-2024} that  Majorana fermions $\psi_{M_{1,2}}=(\psi \pm \psi^{C})/\sqrt{2}$ with  Dirac-type fields $\psi$,  which are the original definition of Majorana \cite{Majorana}, are realized by applying the generalized Pauli-Gursey canonical  transformation to the exact solutions $\psi_{\pm}$ of the type I seesaw model, as shown in \eqref{Pauli--Gursey0}. This formulation is discussed   from a different perspective in the present paper.  The chiral projection $[(1+\gamma_{5})/2]\psi_{M_{1}} $ of a Majorana fermion is not a chiral fermion.
Our formulation using the generalized Pauli-Gursey canonical transformation is based on the uniform charge conjugation and parity operations for all the neutrinos and charged fermions (including quarks), and thus one may feel more confident in the detailed analyses of weak interactions. In the leptogenesis \cite{Fukugita}, for example, one may want to consider the case $\langle \phi \rangle\simeq 0$. Such a case should be understood as the limit  $m_{D}/|m_{R}| \rightarrow {\rm small}$ (or $M_{2}/M_{1} \rightarrow {\rm small}$ in the solution) in \eqref{Seesaw} since $\psi_{+}$ by itself is not identified with a Majorana fermion in the present scheme. 

\section{Conclusion}

A lepton number violating chiral fermion does not necessarily imply a Majorana fermion in $d=4$ as is implied by the theorem on the absence of a Majorana-Weyl fermion in d=4,  but one can  describe well-defined Majorana fermions using an analogue of Bogoliubov canonical transformation \cite{Fujikawa2} in the Type I seesaw model  whose structure is reminiscent of   the BCS theory.

\section*{Acknowledgements}
  
I thank K. Funakubo and A. Tureanu for stimulating discussions. I have benefitted from valuable comments on this subtle issue by H. Sugawara, J. Arafune, K. Hikasa, N. Ohta, and S. Iso.

\end{document}